\documentclass[12pt,preprint]{aastex}

\begin{document}

\title{Rotation and multiple stellar population 
in globular clusters}

\author{Kenji Bekki} 
\affil{
ICRAR,
M468,
The University of Western Australia
35 Stirling Highway, Crawley
Western Australia, 6009
}

\begin{abstract}

We investigate structure and kinematics of the second generation 
of stars (SG) formed from gaseous ejecta of the first generation
of stars (FG) in forming globular clusters (GCs).   
We consider that SG can be formed
from gaseous  ejecta from AGB stars of FG with 
the initial total mass of  $10^6-10^8 {\rm M}_{\odot}$
to explain the present masses of the Galactic GCs. 
Our 3D hydrodynamical simulations with star formation
show that SG formed
in the central regions of FG can have a significant amount
of rotation ($V/\sigma \approx$ $0.8-2.5$).
The rotational amplitude of  SG can depend strongly  on the initial kinematics
of FG.
We thus propose  that  some GCs composed of FG and SG
had a significant
amount of rotation when they were formed. 
We also suggest that although 
later long-term ($\sim 10$ Gyr) dynamical evolution of stars 
can smooth  out the  initial structural and
kinematical differences between FG and SG to a large extent,
initial flattened structures and rotational kinematics
of SG can be imprinted on shapes and internal rotation
of the present  GCs.
We discuss these results in terms of  internal rotation
observed in the Galactic GCs.
\end{abstract}

\keywords{
galaxies:star clusters:general --
galaxies:star clusters:individual: NGC 2808 --
galaxies:star clusters:individual: $\omega$ Centauri --
stars:chemically peculiar 
}

\section{Introduction}

A growing number of
observations have revealed strong evidence for the presence
of multiple stellar populations in some of the Galactic  GCs
(e.g.,  Bedin et al. 2004; Piotto et al. 2005, 2007; 
Lee et al. 2007; Ferraro et al. 2009; Lee et al. 2009; Yong et al. 2009).
The observed Na-O and Mg-Al anti-correlations 
and the possible existence of  very helium-rich stars 
(Y $>0.28$)
in some massive
GCs like NGC 2808 and $\omega$ Cen
are suggested to
be due to the secondary formation of stars (i.e., SG) from
gaseous ejecta from FG that had already formed well
before SG formation (e.g, Carretta et al. 2010).
Many previous works strongly suggested that
the original masses of FG should be  
at least  10 times more  massive 
than the present masses of GCs
in order to explain the observed large mass fraction
of SG (e.g., Bekki \& Norris 2006; Prantzos \& Charbonnel 2006).

If GCs originate from stellar systems
with $M_{\rm s}$ at least larger than $2\times  10^6 {\rm M}_{\odot}$, 
whether they are  very massive super-star clusters 
or dwarf galaxies themselves (e.g., Bekki \& Norris 2006), then
it should be clarified how such massive systems (i.e., FG)
form new stars within them  and evolve with the new stars
(e.g., SG).
D'Ercole et al. (2008) first showed that SG 
can be formed from the gaseous ejecta of AGB stars of
FG with $M_{\rm s}=10^7 {\rm M}_{\odot}$ 
and the large fraction of  FG stars in a forming GC 
can be lost  owing to expansion
and tidal stripping of FG.  
However ,their models are based on one-dimensional hydrodynamical
simulations and have  limitations in predicting
3D structures and kinematics of final stellar systems.
Given that kinematical properties dependent on different
stellar populations have been already derived in $\omega$ Cen
with multiple stellar populations
(e.g., Norris et al. 1997; Sollima et al. 2005)
and will be done in other massive GCs in future,
it is crucial for theoretical studies
to predict  kinematical properties of SG and FG separately.

The purpose of this Letter is  to show that
structural and  kinematical properties of SG can be significantly
different from those of FG 
based on self-consistent 3D hydrodynamical simulations
on star formation within forming GCs.
In particular, we  show that if originally
very massive GCs (FG) have only a small amount of
rotation,  then new stars (SG) formed in the
central regions
can have a significant amount
of rotation depending on initial kinematics of FG.
Based on these results, we  point out that  the origin 
of internal rotation observed in the Galactic GCs
can be closely associated with formation processes of SG in 
original massive stellar systems which  can be  either very massive
super star clusters or dwarf galaxies.

Our future works will investigate  chemical abundances
of SG and long-term evolution of the simulated ``nested structures''
of original GCs.
In the present paper,
we do not discuss the following two key issues
on multiple stellar populations of GCs:
(i) whether massive rotating stars
or AGB ones are responsible for the formation of SG in GCs
and (ii) how chemical evolution
of the clusters proceeds  in their host dwarf galaxies,
both of which have been extensively discussed elsewhere
(e.g., D'Antona et al. 2002; Prantzos \& Charbonnel 2006;
Bekki 2006; Bekki et al. 2007; Renzini 2008; 
Ventura et al. 2008; Carretta et al. 2010).

\section{The model}

We consider that (i) the present GC is composed of FG and SG,
(ii) SG can be formed from gaseous ejecta of AGB stars of FG,
(iii) the initial total mass of FG (i.e., a massive
stellar system) should be much larger than the typical
mass of the Galactic GC ($=2 \times 10^5 {\rm M}_{\odot}$),
and (iv) FG can lose most stars during later dynamical evolution
so that the final mass of a GC 
can be as small as the observed typical mass of the present GCs.
Such initially very massive stellar systems are referred to as
``clusters'' just for convenience, though their stellar masses
are as massive as those of  dwarf galaxies.

We investigate dynamical evolution and star formation of
gaseous ejecta from AGB stars in FG with the total
mass ($M_{\rm s}$) ranging from 
$10^6 {\rm M}_{\odot}$ 
to $10^8 {\rm M}_{\odot}$ by using
the latest version of GRAPE
(GRavity PipE, GRAPE-7) which is the special-purpose
computer for gravitational dynamics (Sugimoto et al. 1990).
We have revised our original GRAPE-SPH code (Bekki 2009)
so that we can investigate 
star formation processes within the above-mentioned 
massive stellar system; the details
of this new code will be described in future papers
(e.g., Bekki 2010).

We adopt the Plummer model with the initial total mass
of $M_{\rm s}$ and the size of $R_{\rm s}$ for FG.
FG is represented by $N=10^5$ stellar particles and
some fraction of the particles are assigned as ``AGB stars''
with initial masses of $m_{\rm s}$
that can eject SPH gas particles with ejection velocities
of $V_{\rm ej}$  with respect to the
centers of the AGB stars. We consider that
gaseous ejecta only from massive AGB stars with
initial masses of $5-8 {\rm M}_{\odot}$ need to be
converted into new stars (i.e., SG) to explain
the chemical abundances of SG (e.g., Renzini 2008).
In the present paper, $V_{\rm ej}$=20 km s$^{-1}$
is adopted, which is a reasonable choice 
for massive AGB stars (e.g., D'Ercole et al. 2008).

The present simulations can not resolve gaseous evolution of each
individual AGB star owing to the adopted numerical resolution
($\sim 0.5$ pc). We therefore assume that each AGB particle initially
has an expanding gaseous sphere which is much larger than the AGB star itself. 
The mass, size, and  temperature of the large gaseous sphere
(represented by SPH particles)  are set to be
$m_{\rm g}$, $r_{\rm g}$ and $T_{\rm g}$, respectively.
We consider that AGB wind can cool down  during expansion
throughout interstellar space (due to radiative cooling)
so that $T_{\rm g}$ can becomes 
much smaller than the original temperature of the wind ($\sim$ 1000K)
when
the size of the  gaseous sphere of the wind become as large as $r_{\rm g}$
(which is $\sim 0.05$ pc). Thus we mainly 
show the results of  the models with  $T_{\rm g}=100$K corresponding
to warm molecular clouds in star-forming regions (e.g., Wilson et al. 1997).

In order to estimate  $m_{\rm g}$ 
corresponds to the total mass ejected from each
AGB star after the main-sequence turn-off, we use
the formula given by Gnedin et al. (2002) to derive
$m_{\rm g}$ as follows: 
\begin{equation}
m_{\rm g}
 =0.916m_{\rm s}-0.444.
\end{equation}
The above equation means that about 83\% of a AGB particle
with $m_{\rm s}=5 {\rm M}_{\odot}$ can be ejected to be
used for further star formation in the present model.
In order to estimate the  mass fraction ($f_{\rm AGB}$)
of AGB progenitor stars with
masses ranging from $5 {\rm M}_{\odot}$
to  $8 {\rm M}_{\odot}$ in FG with the total mass of $M_{\rm s}$,
we assume an
IMF in number defined
as $\psi (m_{\rm I}) = A{m_{\rm I}}^{-s}$,
where $m_{\rm I}$ is the initial mass of
each individual star and the slope $s=2.35$ corresponds to the Salpeter IMF.
The normalization factor $A$ is a function of $M_{\rm s}$,
$m_{\rm l}$ (lower mass cut-off), and $m_{\rm u}$ (upper one).
A is expressed as
$A=\frac{M_{\rm s} \times (2-s)}{{m_{\rm u}}^{2-s}-{m_{\rm l}}^{2-s}}$,
where $m_{\rm l}$ and $m_{\rm u}$ are  set to be   $0.1 {\rm M}_{\odot}$
and  $100 {\rm M}_{\odot}$, respectively.
We adopt the Salpeter IMF and therefore $f_{\rm AGB}=0.042$ in
the present study.

We investigate whether the gas accumulated
in the central regions of FG  
can be sufficient to form new stars by adopting a
simple prescription for star formation.
In the models with ``star formation'',
a gas particle is converted into a collision less new stellar
one if the gas particle meets the following three conditions:
(i) the dynamical time scale of the SPH gas particle
is shorter than the sound crossing time ,  (ii) the gas is converging
(i.e., $\nabla {\bf v} <0$, where ${\bf v}$ is the velocity vector of the
gas particle),
and (iii) the local gas density exceeds densities for
dense cores of molecular clouds ($n > 10^4$ atoms cm$^{-3}$). 
The first  two conditions mimic the Jeans gravitational instability
for gaseous collapse.

The initial stellar system (FG) is assumed to have
a small amount of  rotation with
the ratio of the  initial rotational energy ($T_{\rm rot}$) 
to the total kinetic one ($T_{\rm kin}$) being a free parameter represented
by $s_{\rm rot}$. The parameter values of $s_{\rm rot}$ range
from 0 (no rotation) to 0.2 (rapid rotation). 
The initial rotational velocity of a particle at a distance of $R$ from
the center of FG is $\omega R$, where $\omega$ (constant angular velocity)
 is determined  
such that $s_{\rm rot}$ can be the adopted value.
Therefore, the system has random kinetic energy of 
$(1-s_{\rm rot})T_{\rm kin}$ (due to isotropic velocity dispersion
of stars).

We mainly describe the results of the ``standard'' model with 
$M_{\rm s}=10^7 {\rm M}_{\odot}$,  $R_{\rm s}=100$pc, and $s_{\rm rot}=0.08$,
though we have investigated models with different $M_{\rm s}$, $R_{\rm s}$,
and $s_{\rm rot}$. This is because SG can have 
a total mass ($\sim 2\times 10^5 {\rm M}_{\odot}$)
and a half-mass radius of SG  ($\sim 5$ pc)  that are
similar to those of  the typical Galactic GCs.
The above parameter set is similar to that adopted in
D'Ercole et al. (2008) so that 
the results of our 3D models can be compared with those
of 1D ones by D'Ercole et al. (2008).  
In the present paper, we focus on the models with 
$M_{\rm s}=10^7 {\rm M}_{\odot}$ (with  different $s_{\rm rot}$).
In our future papers (Bekki 2010, in preparation), 
we describe in detail how structures and kinematics of SG depend on 
$M_{\rm s}$, $R_{\rm s}$, $s_{\rm rot}$, and threshold
gas density for star formation.

\section{Results}

Figure 1 shows that new stars (SG) can be formed from gas ejected from
FG in the central region of the original massive cluster 
with  $M_{\rm s}=10^7 {\rm M}_{\odot}$ and $R_{\rm s}=100$pc (T=2.7Myr).
The star formation rate becomes its maximum 
of $\sim 0.25 {\rm M}_{\odot}$ yr$^{-1}$ at T=1.1 Myr then rapidly
declines to be well less than  $0.01 {\rm M}_{\odot}$ yr$^{-1}$
at $T=5.0$ Myr. The starburst with the star formation rate
larger than 0.1 ${\rm M}_{\odot}$ yr$^{-1}$ can last
only for $\sim 1$ Myr.
The half  of the initial gas can be converted
into new stars within the first 3 Myr ($T<3$Myr) and about 61\% of the gas
can be finally converted into the stars within 13.6 Myr.
This short-term starburst during  SG formation
can be seen in all of the models in the
present study.

The half-mass radius ($r_{\rm h}$) of the final cluster is 24.5pc for FG and
6pc for SG, if we consider all particles with $R<250$pc
in the simulation: it should be
noted here that $r_{\rm h}$ can become significantly smaller after
stripping of stars and new stars in later dynamical evolution within
a galaxy hosting the cluster.
The $r_{\rm h}$ ratio of SG to FG is $\sim 0.25$, which means that
SG is significantly more concentrated in the inner region of the cluster.
Owing to gaseous dissipation,  the gas accreting onto the central region
of the cluster can form a flattened and rapidly rotating disk, where
new stars can form if local gas densities
exceeds the adopted threshold gas density
for star formation (=$10^4$ atoms cm$^{-3}$). 
Thus, the new stars (SG) can finally
have a flattened spatial distribution in the central region of the cluster.

Figure  2 shows that there are significant differences in
rotation ($V$) and velocity dispersion ($\sigma$) profiles
in the inner cluster ($R<12$pc)  between FG and SG.
The maximum value of $V$ ($V_{\rm rot}$)
for FG and SG are 3.2 km s$^{-1}$ and 15.3 km s$^{-1}$,
respectively, which clearly demonstrates that the simulated SG is strongly
rotating. 
The maximum value of $\sigma$ (${\sigma}_{\rm 1D}$)
for FG and SG are 16.7 km s$^{-1}$ and 7.9 km s$^{-1}$, respectively,
which means that $V_{\rm rot}/{\sigma}_{\rm 1D}$ for FG and SG
are 0.19 and 1.93, respectively.
The formation of the  dynamically cold and rotating SG is due largely
to gaseous dissipation during gas accretion onto the cluster center.
These results clearly suggest that  initial kinematical
properties can be significantly different between FG and SG in forming GCs.

Figure 3 shows that $V_{\rm rot}$ of SG  strongly depends on $s_{\rm rot}$
of FG
in such a way  that $V_{\rm rot}$ is larger for larger $s_{\rm rot}$ (i.e.,
a larger ratio of rotational energy to total kinetic one for FG).
This result suggests that the original rotational properties of FG in a cluster
can determine rotational amplitude of SG in the cluster.
It is found that $V_{\rm rot}/{\sigma}_{\rm 1D}$ of SG is higher
for larger $s_{\rm rot}$: it is 0.75 for $s_{\rm rot}=0.003$ and
2.53 for $s_{\rm rot}=0.18$.
All of the  four models with different $s_{\rm rot}$ 
show  ${\sigma}_{\rm 1D}$ 
less than 10 km s$^{-1}$, which means that SG is dynamically
cold in forming clusters. 
It should be stressed here that
even in the model with a very small $s_{\rm rot}$ (=0.003),
the final SG shows a significant amount of rotation 
($V_{\rm rot}=7.5$ km s$^{-1}$).  This suggests that only
a small amount of rotation in FG would be necessary to form 
rotating SG in  a forming GC.

The final distributions of new stars (SG) also depend on $s_{\rm rot}$
such that they can be more flattened in models with larger $s_{\rm rot}$.
Figure 4 shows that the final shape of SG in the model 
with $s_{\rm rot}=0.003$ appears to be significantly flattened,
though it is less flattened than the model with $s_{\rm rot}=0.08$ (shown
in Figure 1). This result suggests that only a small amount of rotation
of FG can result in  a significantly flattened distribution of SG.
It is also found that (i)  gas from FG can be more efficiently converted
into new stars (SG) in models with smaller $s_{\rm rot}$ 
and (ii) final distributions of SG are more compact in models
with smaller $s_{\rm rot}$. These results suggest that
internal rotation of FG can play a role in determining structural
and kinematical properties of SG in forming GCs.

\section{Discussion and conclusions}

We have  shown that SG can have significantly  flattened
and compact spatial  distributions and be rotating
in the central regions of FG.
These distinct  nested structures with inner flattened components
with rotation
in original GCs composed of FG and SG  are
not observed in the present  Galactic GCs.
This means that later long-term ($\sim 10$ Gyr) 
dynamical evolution needs to smooth out  the originally 
nested  systems with rotation 
to a large extent.
Recent theoretical study has  shown that the initially nested structures 
of GCs can evolve into structurally homogeneous systems
due to two-body relaxation and tidal force of GCs' host
galaxies within two relaxation timescales (Decressin et al. 2008).
The  FG of GCs can lose most of their stars due to long-term
dynamical relaxation processes and tidal stripping 
whereas SG can remain the same owing to the more compact
distributions: the final (=present)
GCs can be as massive/small as the observed ones (D'Ercole et al. 2008).

Owing to the lack of numerical simulations on dynamical evolution
of nested stellar systems {\it with flattened rotating components},
it is  impossible to discuss the dynamical fate of the simulated
stellar systems in the present study.
A large particle number ($N \sim 10^7$) is required
for investigating this for 
the originally very massive 
systems with $M_{\rm s}\sim 10^7 {\rm M}_{\odot}$
in a fully consistent manner
and thus is currently beyond our reach
(See Vesperini 2010 for a comprehensive review on
short- and long-terms dynamical relaxation processes of evolving GCs).
However it is no doubt that
our future more sophisticated high-resolution
simulations will investigate  whether the simulated nested systems
can finally become those similar to the present GCs both in structures
and kinematics.

Previous observations  investigated structural and kinematical differences
between stellar populations with different chemical abundances 
in the most massive Galactic GC $\omega$ Cen
(e.g., Norris et al. 1997; Ferraro et al. 2002;
Sollima et al. 2005, 2007; Pancino et al. 2007).
For example, Sollima et al. (2005, 2007) found that
(i) stars with possibly large helium abundances (Y$\sim 0.36$)
have a significantly more compact spatial distribution
and (ii) intermediate-metallicity populations are kinematically
cooler than others.
The present study implies that
these structurally and kinematically distinct populations
would be imprint of SG formation processes driven by gas dynamics
in forming GCs. 
Structural and kinematical differences between FG and SG 
for normal GCs in the Galaxy have not been extensively
investigated so far: it is worthwhile for future observations
to confirm whether or not SG in the present GCs show more compact
spatial distributions with a larger amount of rotation
in comparison with those of FG, in particular, for more
massive GCs.

Although evolution of initially flattened clusters into more spherical
present GCs has been discussed in many authors
(e.g., Frenk \& Fall 1982; Einsel \& Spurzem 1999),
the origin of possibly flattened shapes of GCs
{\it at their birth}   have not been
discussed extensively. 
Binary clusters formed from collisions of giant molecular clouds (GMCs)
in the LMC  (Bekki et al. 2004) can finally merge to form flattened
single clusters. 
Bekki and Mackey (2009) recently have shown that merging between
GMCs and star clusters can end up with formation of new stars
with very flattened stellar distributions. 
The present results imply that if  FG 
in original GCs can be lost to a much  larger extent
than SG (with initially very flattened shapes)
owing to more diffuse stellar distributions of FG,
as shown in D'Ercole  et al. (2008),
then the GCs can finally  look like more flattened.

Non-cylindrical differential rotation
with the maximum rotational velocities of 8.0 km s$^{-1}$
at 3-4 core radii for $\omega$ Cen
and 6.5 km s$^{-1}$ at 11-12 core radii for 47 Tuc were revealed
by Meylan \& Mayer (1986).
The internal rotation with 5.7 km s$^{-1}$ at a radius
of 7.5 pc for  47 Tuc was derived by Anderson \& King (2003)
using the {\it Hubble Space Telescope (HST)}.
The Galactic GC M15 is observed to have
an  inner flattened component with possible rotation
(e.g., van den Bosch et al. 2006). 
The observed $V_{\rm rot}/{\sigma}_{\rm 1D}$
ranges from 0.01 to 0.50 for the sampled 12 Galactic GCs,
and
the  mean
$V_{\rm rot}/{\sigma}_{\rm 1D}$
is $\sim 0.2$ (e.g.,  Meylan
\& Heggie 1997).
However it remains observationally unclear whether SG stars in the Galactic GCs
have rotational kinematics.

Recently structural and kinematical differences between 
the blue (bMS) and red (rMS) sequence stars in $\omega$ Cen
have been investigated in detail by a number of authors
(e.g., Sollima et al. 2007; Bellini et al. 2009; 
Anderson \& van der Marel 2010). Given that the bMS stars
are suggested to form from AGB ejecta of rMS ones 
(e.g., Bekki \& Norris 2006),  they correspond to SG in the 
present study  and thus their rotational properties can be discussed 
using the present results on rotational properties of SG.
Although recent observational results
by Anderson \& van der Marel (2010) which show clear internal
rotation in the bMS stars of  $\omega$ Cen
are consistent with the present scenario,
their results on the rotation amplitude of the rMS stars similar
to that of the bMS
can not be simply explained by the present scenario. 
It would be possible that initial kinematical differences
between the bMS and rMS stars 
(i.e., more rotation in the bMS) could have diminished
owing to long-term dynamical evolution within $\omega$ Cen.

Given that the bulk of the GCs' present populations
(typically 67\%) 
are observationally demonstrated to be from  SG 
(e.g., Carretta et al. 2010),
the origin of internal rotation of the GCs can be closely
associated with the formation of rotating SG.
The present simulations have shown  that the rotation of SG
is due largely to dissipative accretion processes
of AGB ejecta from FG with
initially a small amount of rotation.
Therefore  the origin of the observed rotation of GCs 
can be closely associated with rotation in the ``lost generation'' of
stars (i.e., FG)
and thus with rotation of GMCs hosting FG.
We conclude that  shapes, rotational properties, and chemical abundances
of the present GCs all have fossil information on the very
early evolution processes (e.g., star formation
and dynamics of gas from FG) in  forming GCs.

\acknowledgments
We are grateful to the anonymous referee for constructive and
useful comments.
Numerical computations reported here were carried out both on the GRAPE
system
at the University of Western Australia  and on those
at the Center for computational
astrophysics
(CfCA) of the National Astronomical Observatory of Japan.
This work was financially supported by CfCA.

\begin{figure}
\epsscale{.80}
\plotone{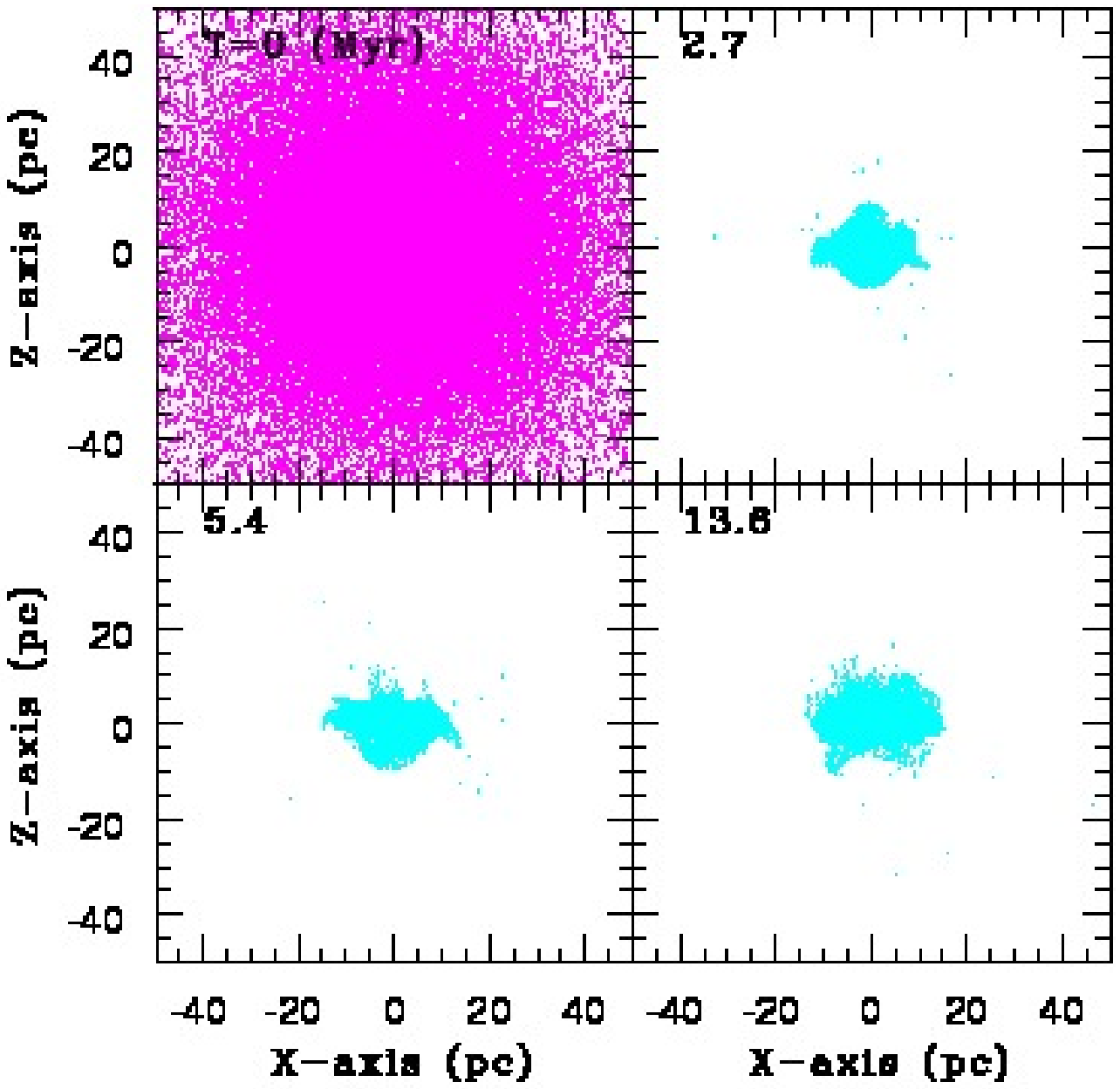}
\figcaption{
Time evolution of the spatial distributions of FG (magenta)
and SG (cyan) in the very massive ``cluster'' with
$M_{\rm s}=10^7 {\rm M}_{\odot}$,  $R_{\rm s}=100$pc,
and $s_{\rm rot}=0.08$. SG is formed as a result of
star formation from gas ejected from AGB stars of FG.
Time (in units of Myr) is shown
in the upper left corner of each panel.
The distribution of FG is shown only for  $T=0$ Myr, because the distribution
can be  almost the same  during 13.6 Myr evolution of the cluster
owing to a smaller amount of total gas mass ejected from AGB stars ($\sim 4$\%)
in FG. 
\label{fig-1}}
\end{figure}

\begin{figure}
\epsscale{.60}
\plotone{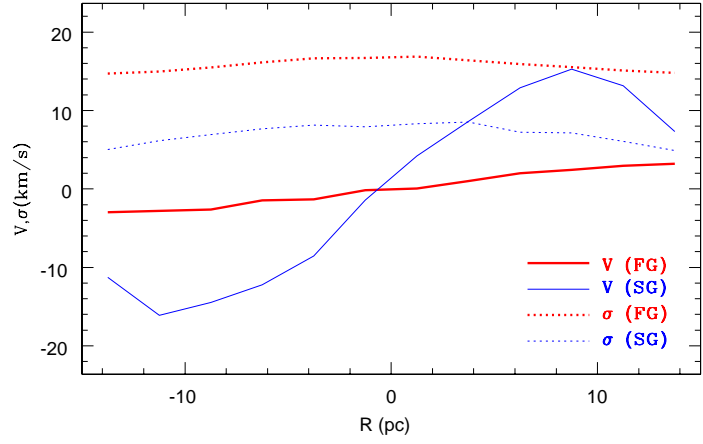}
\figcaption{
The radial profiles of line-of-sight rotational velocity $V$ (solid)
and velocity dispersion $\sigma$ (dotted) for FG (red)
and SG (blue) in the model with
$M_{\rm s}=10^7 {\rm M}_{\odot}$,  $R_{\rm s}=100$pc,
and $s_{\rm rot}=0.08$ at T=13.6 Myr (shown in Figure 1).
\label{fig-2}}
\end{figure}

\newpage

\begin{figure}
\epsscale{.60}
\plotone{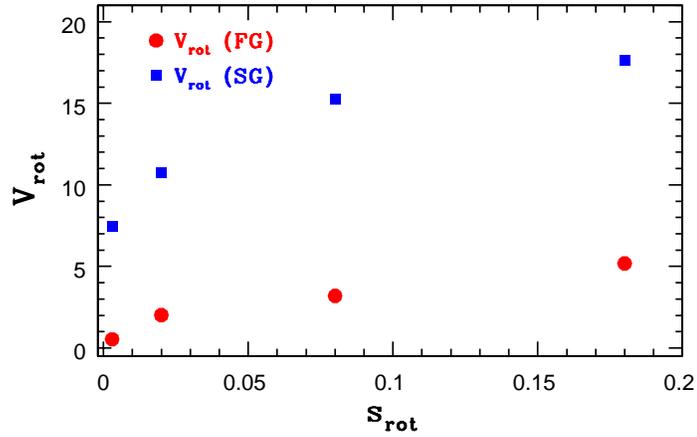}
\figcaption{
The dependences of $V_{\rm rot}$ (maximum $V$) 
for FG (red circles) and SG (blue squares) on
$s_{\rm rot}$ for models with 
$M_{\rm s}=10^7 {\rm M}_{\odot}$ and  $R_{\rm s}=100$pc.
\label{fig-3}}
\end{figure}

\newpage

\begin{figure}
\epsscale{.60}
\plotone{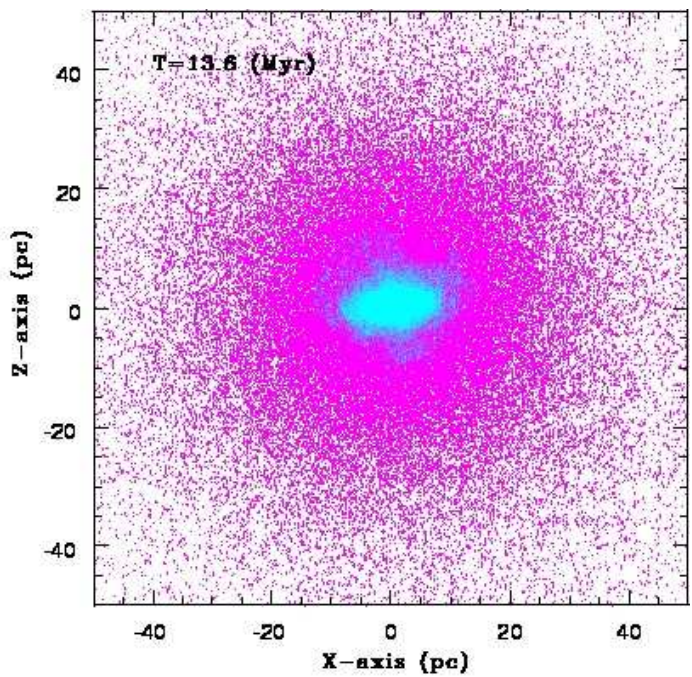}
\figcaption{
Final spatial distributions of FG (magenta)
and SG (cyan) in the model with 
$M_{\rm s}=10^7 {\rm M}_{\odot}$,  $R_{\rm s}=100$pc,
and $s_{\rm rot}=0.003$.
\label{fig-4}}
\end{figure}

\end{document}